# Borophosphene: a New Anisotropic Dirac Cone Monolayer with High Fermi Velocity and Unique Feature of Self-doping

Yang Zhang,[†,‡] Jun Kang,[‡] Fan Zheng,[‡] Peng-Fei Gao,[†] Sheng-Li Zhang,[†] and Lin-Wang Wang*[,‡]

[†]*Ministry of Education Key Laboratory for Nonequilibrium Synthesis and Modulation of Condensed Matter, Department of Applied Physics, School of Science, Xi'an Jiaotong University, Xi'an 710049, China*

[‡]*Materials Science Division, Lawrence Berkeley National Laboratory, Berkeley, CA 94720, USA*

**ABSTRACT:** Two-dimensional (2D) Dirac cone materials exhibit linear energy dispersion at the Fermi level, where the effective masses of carriers are very close to zero and the Fermi velocity is ultrahigh, only 2 ~ 3 orders of magnitude lower than the light velocity. Such the Dirac cone materials have great promise in high-performance electronic devices. Herein, we have employed the genetic algorithms methods combining with first-principles calculations to propose a new 2D anisotropic Dirac cone material, that is, orthorhombic boron phosphide (BP) monolayer named as borophosphene. Molecular dynamics simulation and phonon dispersion have been used to evaluate the dynamic and thermal stability of borophosphene. Because of the unique arrangements of B-B and P-P dimers, the mechanical and electronic properties are highly anisotropic. Of great interest is that the Dirac cone of the borophosphene is robust, independent of in-plane biaxial and uniaxial strains, and can also be observed in its one-dimensional (1D) zigzag nanoribbons and armchair nanotubes. The Fermi velocities are ~ $10^5$ m/s, the same order of magnitude with that of graphene. By using a tight-binding model, the origin of the Dirac cone of borophosphene is analyzed. Moreover, a unique feature of self-doping can be induced by the in-plane biaxial and uniaxial strains of borophosphene and the Curvature effect of nanotubes, which is great beneficial to realizing high speed carriers (holes). Our results suggest that the borophosphene holds a great promise in high-performance electronic devices, which could promote the experimental and theoretical studies to further explore the potential applications of other 2D Dirac cone sheets.

First-principles study.

## 1. INTRODUCTION

Graphene is a typically 2D Dirac cone material with linear energy dispersion at the Fermi level,[1] exhibiting excellent electronic properties such as ballistic charge transport,[2] ultrahigh carrier mobility,[3] and quantum Hall effects.[4] The Fermi velocity of graphene is about $8.3\times10^5$ m/s, only 3 orders of magnitude lower than the light velocity, offering a promising route of developing high-performance electronic devices. Motivated from the fascinating properties and extensive applications presented in graphene, the research passion of others 2D Dirac cone materials has been ignited in the past decade. To date, various types of 2D nanosheets have been successively identified as Dirac cone materials, such as silicene,[5] germanene,[6] carbon allotropes,[7-10] $SiC_3$,[11] $C_4N$,[12] and $Be_3C_2$.[13]

Recently, boron-based Dirac cone materials have attracted great attention because boron atoms have a short covalent radius and flexible chemical bonding that favor the formation of multiple 2D allotropes.[14-24] Using *ab initio* evolutionary structure search, graphene-like ionic boron plane with the *P6/mmm* space group has been predicted to be dynamically and thermally stable, and exhibit double Dirac cones with massless Dirac Fermions.[15] What important and interesting is that its Fermi velocity is as high as $2.3 \times 10^6$ m/s, even one order of magnitude higher than that of graphene. In addition, a Dirac cone can exist in boron $\beta_{12}$ monolayer which has been successfully synthesized on a Ag(111) substrate and exhibits a triangular lattice with arrangements of periodic holes.[21,22] By using high-resolution angle-resolved photoemission spectroscopy, 2D anisotropic Dirac cones have been observed in $\chi_3$ borophene (a monolayer boron sheet),[24] where the carrier mobility is also anisotropic, suggesting great promise in electronic devices with direction-dependent transportation. Recent theoretical studies have confirmed that boron-based 2D compound monolayers can also exhibit the characteristics of Dirac cone,[25,26] suggesting a new route for designing 2D anisotropic Dirac cone materials.

Although the 2D Dirac cone materials provide great potential in electronic devices, there are two issues should be considered: (i) the features of Dirac cone can be maintained in the related 1D systems, including nanoribbons and nanotubes, which is greatly beneficial to constructing the conductive channels or networks utilizing the ultrahigh carrier mobility and reducing the sizes of devices; (ii) charge carriers should be easily controlled for realizing the ultrahigh electron conductivity around the Dirac point. For instance, graphene is a typical semimetal with conically shaped valence and conduction band reminiscent of relativistic Dirac cones for massless particles. A prerequisite for developing graphene based electronics is the reliable control of the type and density of the charge carriers without affecting the conical band structure.[27] Both electron- and hole-doping in graphene can be achieved through surface adsorptions and epitaxial growing graphene on certain substrates,[27-30] which might cause great complexity to implement in device technology. Hence, it is natural to propose that if these two issues mentioned above coexist in one system or can be easily controlled, it will be great benefit to designing the electronic devices with ultrahigh carrier mobility based on 2D Dirac cone materials.

In this study, we have performed the genetic algorithms method combining with first-principles calculations to propose a new 2D anisotropic Dirac cone material of borophosphene. We have found that its Dirac cone is robust, independent of in-plane uniaxial and biaxial strains, and demonstrated to be symmetrically protected using a tight-binding model. What interesting is that such a Dirac cone can also be observed in 1D nanoribbons and nanotubes of borophosphene. Furthermore, a unique feature of self-doping can be induced by in-plane uniaxial and biaxial strains in borophosphene or the Curvature effect in its nanotubes, which is greatly significant for the practical applications of the Dirac cone materials in high-performance electronic devices.

## 2. COMPUTATIONAL DETAILS

All calculations were performed by using first-principles study based on the spin-polarized DFT within the projector augmented wave method,[31,32] as implemented in Vienna *ab initio* simulation package (VASP).[33,34] The generalized gradient approximation (GGA) with the Perdew-Burke-Ernzerhof (PBE) functional was employed to describe the electron exchange-correlation interactions.[35,36] The cut-off of plane-wave kinetic energy and the convergence of total energy were set to be 400 eV and $10^{-5}$ eV. The studied monolayers were modelled in a rectangular cell and located in the *x-y* plane. Because of the application of periodic boundary conditions, a vacuum region over than 15 Å was applied along the *z*-axis to eliminate the interactions between neighbour layers. Thence, the brillouin zone integrations were approximated by 17 × 11 × 1 *k*-point meshes with Gamma centered grid. For the calculations of 1D nanoribbons and nanotubes, 1 × 1 × 17 *k*-point meshes were used. The structural relaxations were performed by computing the Hellmann-Feynman forces using conjugate gradient algorithm within a force convergence of 0.01 eV/Å.[37] In order to explore the dynamic stability, the vibrational properties were investigated using the PHONON package with the forces calculated from VASP software.[38] For the investigations of electronic band structures, the screened hybrid density functional HSE06 was employed,[39] which is considered to be an accurate measure of band gaps for semiconductors.

The global energy minimum configurations of 2D BP monolayers were predicted using our in-house code to implement the genetic algorithms method,[40] which has been demonstrated to be valid in 2D material structure searching.[41] For each generation, the DFT relaxation of the populations was performed by using VASP. The details of our global minimum structure search code were presented in the supporting information (see Figure S1). Herein, two types of stable BP monolayers were identified and shown in Figure S2. One exhibits hexagonal symmetry with the space group of $P\bar{6}m2$ (denoted as *h*-BP monolayer). The other displays orthorhombic symmetry with the space group of *Pmmm* (named as borophosphene). These two sheets are energetically stable and correspond to a minimum in energy surface. For the *h*-BP monolayer, previous studies have confirmed its structural and mechanical stability and semiconducting properties with high carrier mobility.[42-45] However, for the borophosphene, it is a new Dirac cone material in the 2D monolayer families, which is proposed for the first time. Therefore, in this study, we devote to systematically explore the structural stability and electronic properties of the borophosphene.

## 3. RESULTS AND DISCUSSION

Optimized geometrical structure of borophosphene is shown in Figure 1a and the calculated structural parameters are summarized in Table 1. Similar to graphene, the borophosphene is a planar monolayer within hexagonal honeycomb-like structure composed of B-B and P-P dimers. Thus, this layer exhibits orthogonal symmetry with two B and two P atoms in a unit cell. The calculated lattice constants *a* and *b* are 3.22 and 5.57 Å, which are very close to those of *h*-BP monolayer.[42-45] The bond lengths of B-B, P-P and B-P (see Table 1) are comparable to those of $h$-BP monolayer (B-P: 1.86 Å),[45] borophene (B-B: 1.62 Å),[46] and phosphorene (P-P: 2.22 Å),[47,48] indicating that the chemical bonds in the borophosphene are strong. Cohesive energy is one of the key factors to evaluate the feasibility of experimental synthesis for the predicted 2D materials,[49-51] which is calculated according to the formula of $E_c = (E_B + E_P - E_{BP})/2$, where $E_B$, $E_P$, and $E_{BP}$ are the total energies of a single B atom, a single P atom, and one BP molecule in the borophosphene. The resultant cohesive energy is 4.82 eV/atom, a little smaller than the $h$-BP monolayer (4.99 eV/atom) and much higher than these synthesized black and blue phosphorenes (both are ~ 3.48 eV/atom). Our results suggest that the borophosphene holds a high possibility to be synthesized under certain experimental conditions.

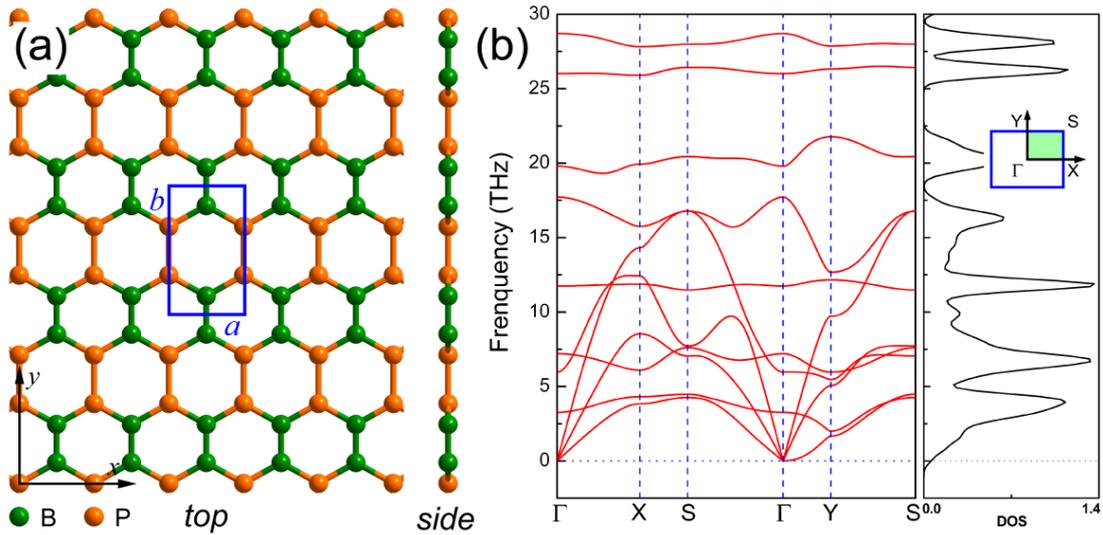

**Figure 1.** (a) Optimized geometrical structure viewed from different directions and (b) phonon spectrum together with density of state (DOS) of borophosphene. Green and orange balls denote B and P atoms.

In order to evaluate the stability of borophosphene, phonon spectrum with density of state (DOS) is calculated and shown in Figure 1b. Obviously, there are no imaginary modes of lattice vibrations in the whole Brillouin zone, suggesting that the borophosphene is dynamically stable. Further examination of the thermal stability is performed by using *ab initio* molecular dynamics (MD) simulations. A large 6 × 4 supercell is adopted by heating the structure to 300 and 600 K within a canonical ensemble, which corresponds to the sizes of 19.33 and 22.26 Å. In each case, the simulation is last for 10 ps with a time step of 1.5 fs. At the end of each simulation, the final structure is carefully examined. The snapshots viewed from different directions are presented in Figure S3. At room temperature, the borophosphene can maintain its planar structure very well.

Even at a high temperature of 600 K, it can still withstand slight distortions which are not sufficient to destroy the B-B, P-P and B-P chemical bonds. In a word, the results of phonon spectrum calculations and MD simulations suggest the highly dynamic and thermal stability of borophosphene. Different from others hexagonal honeycomb sheets, the unique arrangements of B-B and P-P dimers in the borophosphene could contribute to anisotropic properties, including mechanical and electronic ones.

**Table 1.** Properties of 2D BP monolayers: plane group, lattice constant $a$ & $b$ (Å), bond length (Å), cohesive energy $E_c$ (eV/atom), and band gap $E_g$ (eV) predicted by using PBE functional and HSE06 method.

| Models | Plane group | lattice constant | Bond length | | | $E_c$ | $E_g$ |
| --- | --- | --- | --- | --- | --- | --- | --- |
| | | | B-B | B-P | P-P | | |
| $h$-BP | $P\bar{6}m2$ | $a = b = 3.21$ | - | 1.86 | - | 4.99 | 0.91 (PBE) <br> 1.37 (HSE06) |
| Borophosphene | $Pmmm$ | $a = 3.22$ <br> $b = 5.57$ | 1.65 | 1.85 | 2.11 | 4.82 | semimetal |

The mechanical stability of borophosphene is examined by calculating the linear elastic constants. The calculated results of 2D linear elastic constants are $C_{11}= 154.7$ N/m, $C_{22}= 136.6$ N/m, $C_{12}= 34.8$ N/m, and $C_{44}= 49.6$ N/m. For an orthorhombic 2D monolayer, the stability criteria are $C_{11} > 0$, $C_{22} > 0$, $C_{44} > 0$, and $C_{11}C_{22} > C_{12}^2$.[52] Obviously, all the elastic constants satisfy those conditions above, confirming that the borophosphene is mechanically stable. Subsequently, the in-plane Young's modulus and Poisson ratio along an arbitrary direction $\theta$ ($\theta$ is the angle relative to the positive $x$ direction of the sheet) can be expressed as[53]

$$E(\theta) = \frac{C_{11}C_{22} - C_{12}^2}{C_{11}s^4 + C_{22}c^4 + \left(\frac{C_{11}C_{22} - C_{12}^2}{C_{44}} - 2C_{12}\right)c^2s^2},$$

$$\nu(\theta) = -\frac{\left(C_{11} + C_{22} - \frac{C_{11}C_{22} - C_{12}^2}{C_{44}}\right)c^2s^2 - C_{12}(c^4 + s^4)}{C_{11}s^4 + C_{22}c^4 + \left(\frac{C_{11}C_{22} - C_{12}^2}{C_{44}} - 2C_{12}\right)c^2s^2},$$

where $c = \cos(\theta)$ and $s = \sin(\theta)$. The polar diagrams of $E(\theta)$ and $\nu(\theta)$ for the borophosphene are displayed in Figure 2. It can be found that both the in-plane Young's modulus and the Poisson ratio are highly anisotropic. The Young's modulus ranges from 126.2 to 145.8 N/m. A maximum value is observed along the zigzag direction (i.e., *x* axis denoted in Figure 1a), while a minimum one is presented at the angles of 60° and 120° to the *x* axis. In comparison with others typical 2D nanosheets, such as graphene (~ 340 ± 40 N/m),[54] BC$_3$ (~ 316 N/m),[55] and BN (~ 267 N/m),[56] the in-plane Young's modulus of borophosphene is much smaller, indicating a much softer layer. Even so, the borophosphene is still much harder than the experimentally synthesized silicene (~ 62 N/m),[57] which verifies the strong bonding in such a planar sheet.

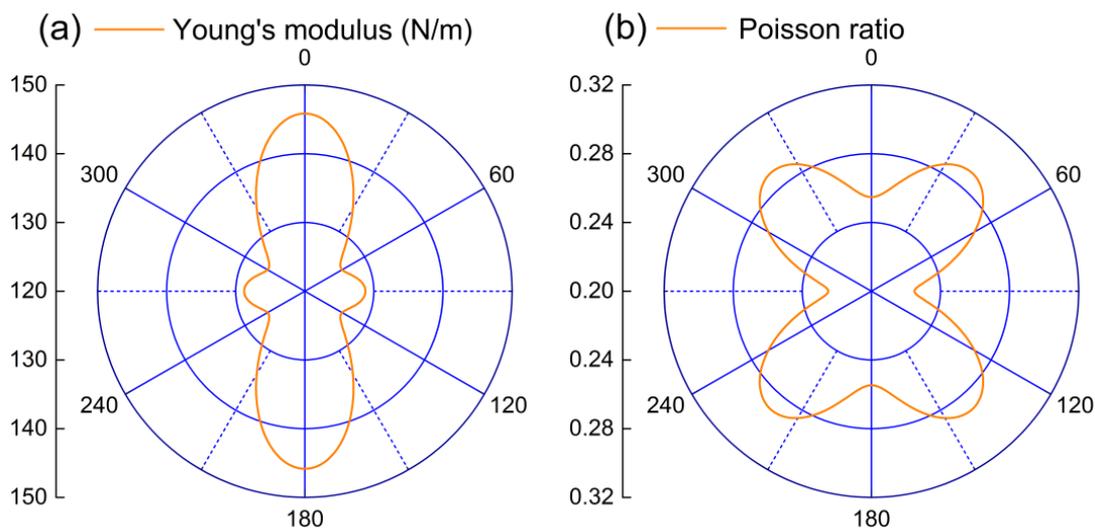

**Figure 2.** Polar diagrams of (a) the in-plane Young's modulus and (b) the Poisson ratio for the borophosphene.

The most intriguing aspect of the borophosphene lies in its electronic property. As shown in Figure 3, the electronic band structure calculated by using the PBE functional demonstrates the borophosphene is significantly anisotropic in the electronic properties. Along the Γ-X direction (i.e., *x* axis denoted in Figure 1a), two energy bands intersect linearly at the Fermi level, forming a zero-gap Dirac cone at the *k*-point of D = π/*a*·(0.75, 0, 0). Whereas a wide band gap of 2.79 eV is observed along the Γ-Y direction (i.e., *y* axis denoted in Figure 1a). Further confirmation of the Dirac cone feature is done by employing a more accurate method of HSE06 functional. Along the Γ-Y direction, a much larger band gap of 3.51 eV is presented, exhibiting poor electronic conductivity along the armchair direction. Moreover, we explore the effect of spin-orbit coupling (SOC) on the band structure of the borophosphene. The results show that no observable band gap occurs because both B and P are light elements and the SOC effect is rather weak. In view of the PDOS (see Figure 3 and Figure S4) and spatial charge density (see the inset of Figure 3 and Figure S5), one can find that the bands around the Dirac cone are occupied by only the $p_z$ orbitals from both the B and P atoms, forming π and π* bonds. The orbitals of *s*, $p_x$, and $p_y$ contribute to the deep bands with the hybridization of σ bonds.

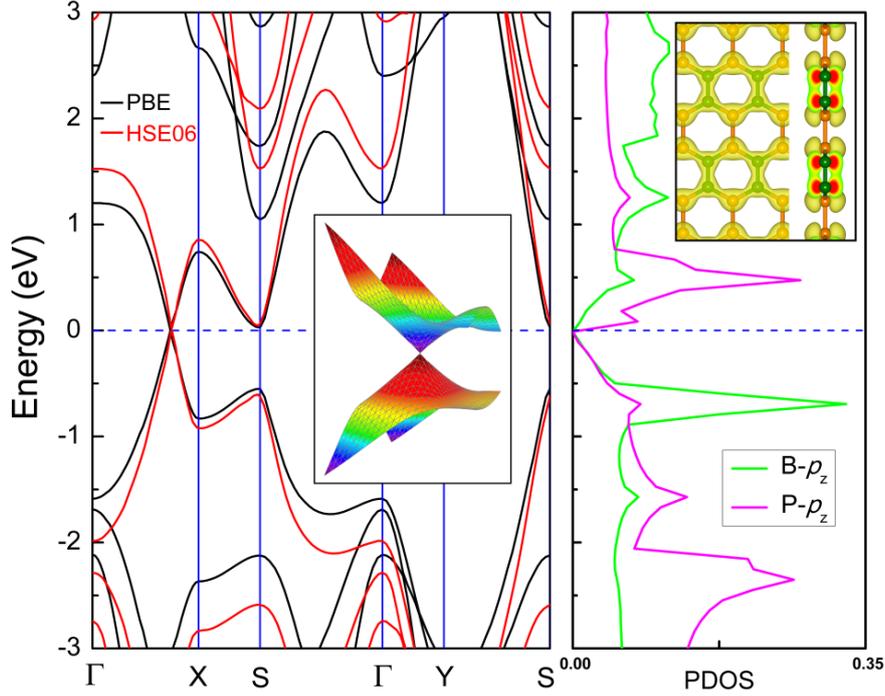

**Figure 3.** Electronic band structure of borophosphene calculated by using the PBE (black lines) and HSE06 (red lines) functional and projected density of state (PDOS). Note that only the $p_z$ orbital contributes to the Dirac cone and is shown here, others are displayed in Figure S4. The insets are 3D electronic band structure and spatial charge density of the Dirac Cone. The Fermi level is set to zero and denoted by blue dash lines.

In Dirac cone materials, the Fermi velocity is the most important feature, because the linear dispersion of energy bands suggests the zero effective mass of the carriers near the Fermi level. Besides, different slopes along the $kx$ and $ky$ directions around the Dirac cone means that the velocities of Dirac Fermions are distinct and strongly depend on the directions. In order to explore it, we calculate three-dimensional (3D) Dirac cone of the borophosphene and extract 2D band structures along the $kx$ and $ky$ directions. The results are shown in Figure S6. It is clear to see that the valence and the conduction bands in the vicinity of the Dirac point suggest the presence of a distorted Dirac cone. The slopes of the bands along the $kx$ and $ky$ directions are ± 28.5 and ± 19.5 eV·Å, implying direction-dependent Fermi velocities in the borophosphene. By fitting linearly the energy bands around the Dirac point, the Fermi velocity of the Dirac Fermions can be obtained according to the equation of $\left(v_f = \frac{1}{\hbar}\frac{\partial E}{\partial k}\right)$. It is found that, by using the PBE functional, the Fermi velocities are predicted to be $6.51 \times 10^5$ and $4.71 \times 10^5$ m/s along the $kx$ and $ky$ directions, which are in the same order of magnitude with that of graphene (8.2 × 10⁵ m/s in ref. 7). A little bit larger Fermi velocity of 7.60 × 10⁵ m/s is evaluated along the *kx* direction by using the HSE06 method. In order to confirm the reliability of our results, we use the same computational method with the PBE functional to calculate the Fermi velocity of graphene. The result of 8.24×10⁵ m/s is in good agreement with the previous study.[7] Hence, one can conclude that the borophosphene can be expected to have high carrier mobility, which is of great benefit to the future electronics.

To further explore the origin of the Dirac cone of borophosphene, we construct a tight-binding model with only $p_z$ orbitals as basis, namely:

$$\widehat{H} = \sum_i \varepsilon_i a_i^+ a_i + \sum_{ij} t_{ij} a_i^+ a_j.$$

Here $i$ and $j$ indicate different atoms in the unit cell. $a^+$ and $a$ are creation and annihilation operators, respectively. $\varepsilon_i$ is the on-site energy for atom $i$, and $t_{ij}$ is the hopping term between atoms $i$ and $j$. Only the hopping between the nearest-neighbor is considered in this model. Due to the symmetry of the structure, there are only two types of $\varepsilon_i$ ($\varepsilon_B$ and $\varepsilon_P$) and three types of $t_{ij}$ ($t_{B-B}$, $t_{P-P}$, and $t_{B-P}$). The Hamiltonian is a 4 × 4 matrix:

$$\langle i | H(\boldsymbol{k}) | j \rangle = \begin{pmatrix} E_0 + E_d/2 & t_{B-P}(1+e^{-ik_x a}) & 0 & t_{P-P}e^{ik_y b} \\ t_{B-P}(1+e^{ik_x a}) & E_0 - E_d/2 & t_{B-B} & 0 \\ 0 & t_{B-B} & E_0 - E_d/2 & t_{B-P}(1+e^{ik_x a}) \\ t_{P-P}e^{-ik_y b} & 0 & t_{B-P}(1+e^{-ik_x a}) & E_0 + E_d/2 \end{pmatrix},$$

where $a$ and $b$ are the lattice constants, $E_0=(\varepsilon_P+\varepsilon_B)/2$, and $E_d=\varepsilon_P-\varepsilon_B$. The form of the matrix does not change as long as the orthogonal symmetry of the structure is preserved. Along the Γ-X high-symmetry line, $k_y = 0$. In this case, one can diagonalize the Hamiltonian and obtain the following four bands:

$$E_1(k_x) = E_0 - \frac{t_{P-P}+t_{B-B}}{2} - \frac{\sqrt{(E_d - t_{P-P}+t_{B-B})^2 + 8t_{B-P}^2[1+\cos(k_x a)]}}{2},$$

$$E_2(k_x) = E_0 - \frac{t_{P-P}+t_{B-B}}{2} + \frac{\sqrt{(E_d - t_{P-P}+t_{B-B})^2 + 8t_{B-P}^2[1+\cos(k_x a)]}}{2},$$

$$E_3(k_x) = E_0 + \frac{t_{P-P}+t_{B-B}}{2} - \frac{\sqrt{(E_d + t_{P-P}-t_{B-B})^2 + 8t_{B-P}^2[1+\cos(k_x a)]}}{2},$$

$$E_4(k_x) = E_0 + \frac{t_{P-P}+t_{B-B}}{2} + \frac{\sqrt{(E_d + t_{P-P}-t_{B-B})^2 + 8t_{B-P}^2[1+\cos(k_x a)]}}{2}.$$

Note that the bands $E_2$ and $E_3$ are not coupled to each other. Thus, as long as there is a solution for $E_2(k_x)=E_3(k_x)$, then these two bands will cross along the Γ-X path, leading to the formation of the Dirac cone. If we assume that the magnitudes of all the $t_{ij}$ are similar ($t_{B-B} \approx t_{P-P} \approx t_{B-P} \approx t$), then the existence of the solution of $E_2(k_x)=E_3(k_x)$ requires $-1 < \cos(k_x a) = (-4t^2 - E_d^2)/8t^2 < 1$. The upper bound is always satisfied, and the lower bound requires $|E_d|<2t$. Therefore, as long as $|E_d|<2t$, there will be Dirac cones in the borophosphene. Then we fit our tight-binding model to DFT calculated band structure and the results are shown in Figure 4. It is seen that the tight-binding bands near the Fermi level agrees well with the DFT calculated ones. The corresponding parameters are found to be $\varepsilon_B$=-0.812 eV, $\varepsilon_P$=0.917 eV, $t_{B-B}$=1.555 eV, $t_{P-P}$=1.749 eV, and $t_{B-P}$=1.591 eV. Note that the magnitudes of all the $t_{ij}$ are indeed similar ($t$ is around 1.6 eV). The $E_d$ is calculated to be 1.729 eV, which is smaller than 2$t$ (~3.2 eV). As a result, the Dirac cone appears. Moreover, when an external strain is applied, the above conclusion holds true if the orthogonal symmetry is not broken by the strain (like the case of in-plane biaxial or uniaxial strain), since the form of the Hamiltonian is not changed. However, if the strain breaks this symmetry (e.g., an in-plane shear strain), then the $E_2$ and $E_3$ band could couple with each other, leading to a band anti-crossing and opens a band gap.

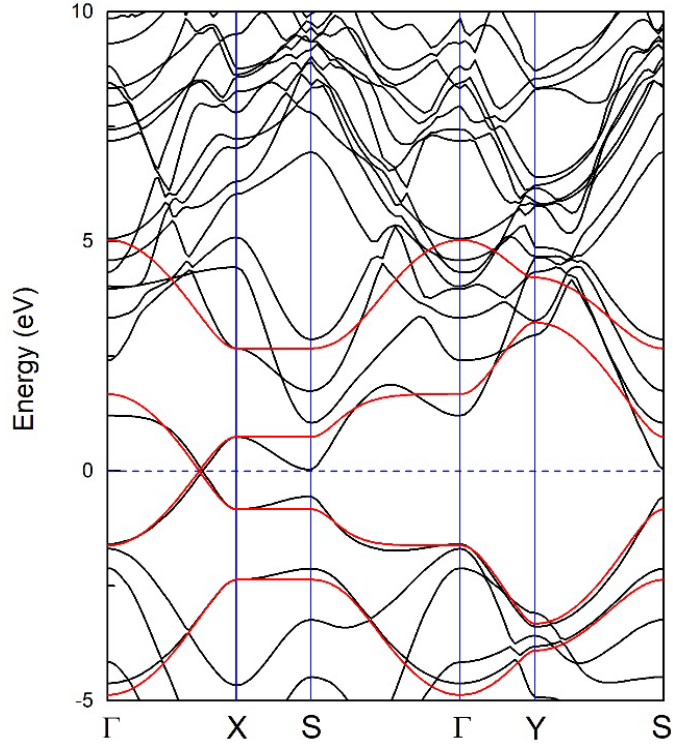

**Figure 4.** Band structures calculated by using a tight-binding model (red lines) and DFT with PBE functional (black lines).

For examining the symmetrical protection of Dirac cone, in-plane biaxial and uniaxial strains are applied to investigate the effects on the band structure of borophosphene. The corresponding results are shown in Figure 5. Obviously, the Dirac cone can still be preserved, independent of the biaxial or uniaxial strain. However, an in-plane shear strain can easily open a band gap (see Figure S7). Our DFT calculations demonstrate the reliability of the tight-binding model and the symmetry protection of the Dirac cone in borophosphene.

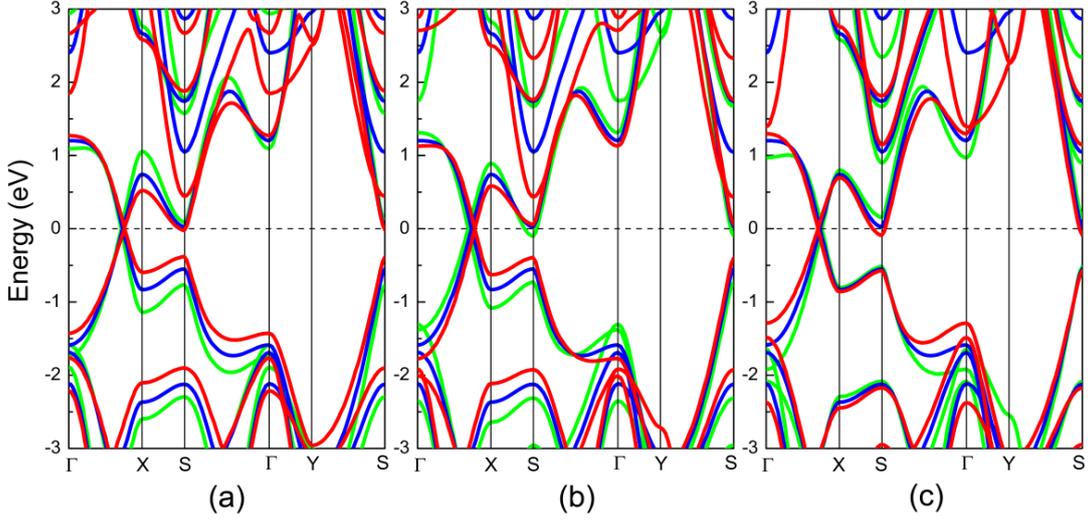

**Figure 5.** Variations of electronic band structure against different in-plane strains: (a) biaxial and uniaxial strains along (b) armchair (*y* axis) and (c) zigzag (*x* axis) directions. Green lines: -5 %, blue ones: 0 %, and red ones: 5 %.

Interestingly, the in-plane strains can induce the lowest conduction band at the *S* point declining and then crossing the Fermi level, forming a unique feature of self-doping in the borophosphene. In a Dirac cone material, the feature of the self-doping is of great significant to realize ultra-high carrier mobility because the Fermi level will be slightly lower/higher than the Dirac point where the electro state is zero, rendering a p-type/n-type doping. In graphene, such the self-doping feature cannot be induced by the in-plane biaxial and uniaxial strains.[58,59] As shown in Figure 5b and 5c, the self-doping feature is very sensitive to the in-plane strains. A tension strain of the zigzag direction (i.e., *x* axis) or a compression strain of the armchair direction (i.e., *y* axis) can result in the feature of self-doping and the borophosphene is metallic. On the contrary, the opposite strain can cause the self-doping to disappear and the borophosphene is a zero-gap semimetal. The Fermi velocity of the borophosphene is not strongly dependent on the in-plane strains. For instances, the biaxial strain can slightly increase the Fermi velocity from $6.51 \times 10^5$ m/s (neutral) to $6.53 \times 10^5$ m/s (5 %) and $7.11 \times 10^5$ m/s (- 5 %). For the uniaxial strain, a maximum value of $6.88 \times 10^5$ m/s can be obtained when an armchair compressive strain of - 5 % is applied, while a minimum one of $6.42 \times 10^5$ m/s can be found at an armchair tensile strain of 5 %. All these velocities are in the same order of magnitude and comparable of that of graphene, suggesting that the in-plane biaxial and uniaxial strains have no significant influence on the Fermi velocity.

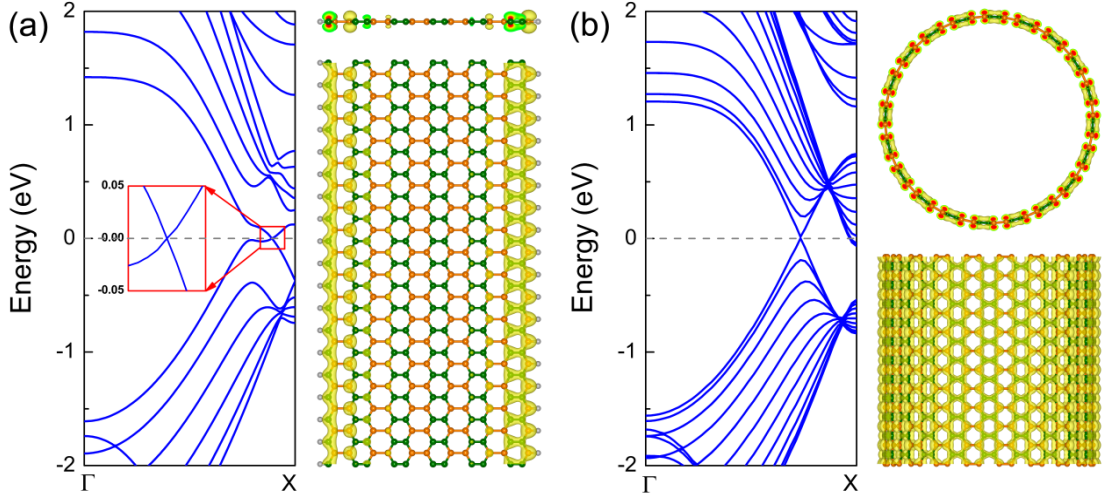

**Figure 6.** Electronic band structures of 1D borophosphene (a) zigzag nanoribbon ($N = 11$) and (b) armchair nanotube (20, 20). The size of nanoribbon is defined in Figure S8. The spatial charge density around the Dirac cone is displayed in the right panel.

Generally, when cutting or rolling 2D Dirac materials into 1D nanoribbons or nanotubes, the Dirac cone will disappear due to the effects of quantum confinement or curvature. However, for the borophosphene, it is completely different. As shown in Figure 6, the features of Dirac cone can still be observed in the 1D zigzag nanoribbons and armchair nanotubes. Whether the nanoribbons are Dirac cone materials, it depends on their edges. Here, we consider the nanoribbons with symmetrical and asymmetrical edges (see Figure S8). To eliminate the effects of dangling bonds, the edges are saturated by hydrogen atoms. It is found that the nanoribbons with asymmetrical edges present the Dirac cones with a nearly constant Fermi velocity of ~ $3.65 \times 10^5$ m/s, independent of the width sizes. For the nanotubes, the armchair ones are Dirac cone materials (see Figure S9), while the zigzag ones are not. Similar to the nanoribbons, the Fermi velocity of armchair nanotubes remains approximately a constant value of ~ $6.62 \times 10^5$ m/s, regardless of the curvature effects. For both the zigzag nanoribbon and armchair nanotube, their high velocities of Dirac Fermions are attributed to the $p_z$ orbitals of B and P atoms. The former mainly comes from the edge states, while the latter originates from the whole surface.

## 4. CONCLUSION

Using the genetic algorithms methods combining with first-principles calculations, we have proposed a new 2D anisotropic Dirac cone material of borophosphene, which has been demonstrated to be dynamically, thermally and mechanically stable. Because of the unique arrangements of B-B and P-P dimers, its mechanical and electronic properties are highly anisotropic. Intriguingly, the Dirac cone of the borophosphene is robust in spite of in-plane biaxial and uniaxial strains, and can also be preserved in 1D zigzag nanoribbons and armchair nanotubes. Such the robust Dirac cone in the borophosphene is attributed to the $p_z$ orbitals of B and P atoms, and further analyzed to be symmetrically protected by using a tight-binding model. The Fermi velocity (~ $10^5$ m/s) is in the same order of magnitude with that of graphene. In addition, the in-plane strains of the borophosphene and the Curvature effect of the armchair nanotube can easily induce a unique feature of self-doping, which is great beneficial to realizing the high speed carriers (holes) in the Dirac cone materials. Our results of Dirac cone and self-doping features suggest that the ultrahigh carrier mobility can be achieved in borophosphene-based electronic devices.

## ASSOCIATED CONTENT

**Supporting Information**

Detailed description of the global energy minimum structure search code and structural prediction, dynamical stability, PDOS, charge density state, and electronic band structures.

## AUTHOR INFORMATION

**Corresponding Author**

*E-mail: lwwang@lbl.gov (L.-W. Wang)

      yzhang520@mail.xjtu.edu.cn (Y. Zhang)

**Notes**

The authors declare no competing financial interest.

## ACKNOWLEDGMENTS

This work is supported by the National Natural Science Foundation of China (Grant No. 11204232 and 11604254), the Natural Science Fundamental Research Program of Shaanxi Province of China (Grant No. 2019JM-190), and Special Guidance Funds for the Construction of World-class Universities (Disciplines) and Characteristic Development in Central Universities in China. The authors acknowledge the computational support from the HPC Platform, Xi'an Jiaotong University.

**REFERENCES**

(1) Geim, A. K.; Novoselov, K. S. The Rise of Graphene. *Nat. Mater.* **2007**, *6*, 183.

(2) Bliokh, Y. P.; Freilikher, V.; Nori, F. Ballistic charge transport in graphene and light propagation in periodic dielectric structures with metamaterials: A comparative study. *Phys. Rev. B: Condens. Matter Mater. Phys.* **2013**, *87*, 245134.

(3) Castro Neto, A. H.; Guinea, F.; Peres, N. M. R.; Novoselov, K. S.; Geim, A. K. The electronic properties of grapheme. *Rev. Mod. Phys.* **2009**, *81*, 109.

(4) Zhang, Y.; Tan, Y. W.; Stormer, H. L.; Kim, P. Experimental observation of the quantum Hall effect and Berry's phase in graphene. *Nature* **2005**, *438*, 201.

(5) Jose, D.; Datta, A. Understanding of the Buckling Distortions in Silicene. *J. Phys. Chem. C* **2012**, *116*, 24639.

(6) Cahangirov, S.; Topsakal, M.; Aktürk, E.; Şahin, H.; Ciraci, S. Two- and One-Dimensional Honeycomb Structures of Silicon and Germanium. *Phys. Rev. Lett.* **2009**, *102*, 236804.

(7) Malko, D.; Neiss, C.; Viñes, F.; Görling, A. Competition for Graphene: Graphynes with Direction-Dependent Dirac Cones. *Phys. Rev. Lett.* **2012**, *108*, 086804.

(8) Chen, J.; Xi, J.; Wang, D.; Shuai, Z. Carrier Mobility in Graphyne Should Be Even Larger Than That in Graphene: A Theoretical Prediction. *J. Phys. Chem. Lett.* **2013**, *4*, 1443.

(9) Xu, L. C.; Wang, R. Z.; Miao, M. S.; Wei, X. L.; Chen, Y. P.; Yan, H.; Lau, W. M.; Liu, L. M.; Ma, Y. M. Two Dimensional Dirac Carbon Allotropes from Graphene. *Nanoscale* **2014**, *6*, 1113.

(10) Wang, Z.; Zhou, X. F.; Zhang, X.; Zhu, Q.; Dong, H.; Zhao, M.; Oganov, A. R. Phagraphene: A Low-Energy Graphene Allotrope Composed of 5-6-7 Carbon Rings with Distorted Dirac Cones. *Nano Lett.* **2015**, *15*, 6182.

(11) Qin, X.; Wu, Y.; Liu, Y.; Chi, B.; Li, X.; Wang, Y.; Zhao, X. Origins of Dirac Cone Formation in $AB_3$ and $A_3B$ (A, B= C, Si, and Ge) Binary Monolayers. *Sci. Rep.* **2017**, *7*, 10546.

(12) Pu, C.; Zhou, D.; Li, Y.; Liu, H.; Chen, Z.; Wang, Y.; Ma, Y. Two-Dimensional $C_4N$ Global

Minima: Unique Structural Topologies and Nanoelectronic Properties. *J. Phys. Chem. C* **2017**, *121*, 2669.

(13) Wang, B.; Yuan, S.; Li, Y.; Shi, L.; Wang, J. A New Dirac Cone Material: A Graphene-Like Be$_3$C$_2$ Monolayer. *Nanoscale* **2017**, *9*, 5577.

(14) Zhou, X. F.; Dong, X.; Oganov, A. R.; Zhu, Q.; Tian, Y. J.; Wang, H.-T. Semimetallic Two-Dimensional Boron Allotrope with Massless Dirac Fermions. *Phys. Rev. Lett.* **2014**, *112*, 0855502.

(15) Ma, F. X.; Jiao, Y. L.; Gao, G. P.; Gu, Y. T.; Bilic, A.; Chen, Z. F.; Du A. J. Graphene-like Two-Dimensional Ionic Boron with Double Dirac Cones at Ambient Condition. *Nano Lett.* **2016**, *16*, 3022.

(16) Zhang, H.; Xie, Y.; Zhang, Z.; Zhong, C.; Li, Y.; Chen, Z.; Chen, Y. Dirac Nodal Lines and Tilted Semi-Dirac Cones Coexisting in a Striped Boron Sheet. *J. Phys. Chem. Lett.* **2017**, *8*, 1707.

(17) Yi, W.-C.; Liu, W.; Botana, J.; Zhao, L.; Liu, Z.; Liu, J.-Y.; Miao M.-S. Honeycomb Boron Allotropes with Dirac Cones: A True Analogue to Graphene. *J. Phys. Chem. Lett.* **2017**, *8*, 2647.

(18) Gupta, S.; Kutana, A.; Yakobson B. I. Dirac Cones and Nodal Line in Borophene. *J. Phys. Chem. Lett.* **2018**, *9*, 2757.

(19) Kou, L. Z.; Ma, Y. D.; Tang, C.; Sun Z. Q.; Du, A. J.; Chen, C. F. Auxetic and Ferroelastic Borophane: A Novel 2D Material with Negative Possion's Ratio and Switchable Dirac Transport Channels. *Nano Lett.* **2016**, *16*, 7910.

(20) Xu, L.-C.; Du, A. J.; Kou, L. Z. Hydrogenated Borophene as a Stable Two-dimensional Dirac Material with an Ultrahigh Fermi Velocity. *Phys. Chem. Chem. Phys.* **2016**, *18*, 27284.

(21) Feng, B.; Zhang, J.; Zhong, Q.; Li, W.; Li, S.; Li, H.; Cheng, P.; Meng, S.; Chen, L.; Wu, K. Experimental Realization of Two-Dimensional Boron Sheets. *Nat. Chem.* **2016**, *8*, 563.

(22) Feng, B.; Zhang, J.; Liu, R.-Y.; Iimori, T.; Lian, C.; Li, H.; Chen, L.; Wu, K.; Meng, S.; Komori, F.; Matsuda, I. Direct Evidence of Metallic Bands in a Monolayer Boron Sheet. *Phys. Rev. B: Condens. Matter Mater. Phys.* **2016**, *94*, 041408.

(23) Feng, B.; Sugino, O.; Liu, R.-Y.; Zhang, J.; Yukawa, R.; Kawamura, M.; Iimori, T.; Kim, H.; Hasegawa, Y.; Li, H.; Chen, L.; Wu, K.; Kumigashira, H.; Komori F. Dirac Fermions in Borophene. *Phys. Rev. Lett.* **2017**, *118*, 096401.

(24) Feng, B.; Zhang, J.; Ito, S.; Arita, M.; Cheng, C.; Chen, L.; Wu, K.; Komori, F.; Sugino, O.;

Miyamoto, K. Discovery of 2d Anisotropic Dirac Cones. *Adv. Mater.* **2018**, *30*, 1704025.

(25) Zhang, H. J.; Li, Y. F.; Hou, J. H.; Du, A. J.; Chen, Z. F. Dirac State in the FeB$_2$ Monolayer with Graphene-Like Boron Sheet. *Nano Lett.* **2016**, *16*, 6124.

(26) Zhao, Y.; Li, X. Y.; Liu, J. Y.; Zhang, C. Z.; Wang Q. A New Anisotropic Dirac Cone Material: A B$_2$S Honeycomb Monolayer. *J. Phys. Chem. Lett.* **2018**, *9*, 1815.

(27) Gierz, I.; Riedl, C.; Starke, U.; Ast, C. R.; Kern, K. Atomic Hole Doping of Graphene. *Nano Lett.* **2008**, *8*, 4603.

(28) Chen, W.; Chen, S.; Qui, D. C.; Gao, X. Y.; Wee, A. T. S. Surface Transfer p-Type Doping of Epitaxial Graphene. *J. Am. Chem. Soc.* **2007**, *129*, 10418.

(29) Wehling, T. O.; Novoselov, K. S.; Morozov, S. V.; Vdovin, E. E.; Katsnelson, M. I.; Geim, A. K.; Lichtenstein, A. I. Molecular Doping of Graphene. *Nano Lett.* **2008**, *8*, 173.

(30) Liu, H.; Liu, Y.; Zhu, D. Chemical doping of grapheme. *J. Mater. Chem.* **2011**, *21*, 3335.

(31) Blöchl, P. E. Projector augmented-wave method. *Phys. Rev. B: Condens. Matter Mater. Phys.* **1994**, *50*, 17953.

(32) Kresse, G.; Joubert, D. From ultrasoft pseudopotentials to the projector augmented-wave method. *Phys. Rev. B: Condens. Matter Mater. Phys.* **1999**, *59*, 1758.

(33) Kresse, G.; Furthmüller, J. Efficient iterative schemes for ab initio total-energy calculations using a plane-wave basis set. *Phys. Rev. B: Condens. Matter Mater. Phys.* **1996**, *54*, 11169.

(34) Kresse G.; Furthmüller, J. Efficiency of ab-initio total energy calculations for metals and semiconductors using a plane-wave basis set. *Comp. Mater. Sci.* **1996**, *6*, 15.

(35) Perdew, J. P.; Burke, K.; Ernzerhof, M. Generalized Gradient Approximation Made Simple. *Phys. Rev. Lett.* **1996**, *77*, 3865.

(36) Zhang, Y.; Yang, W. Comment on "Generalized Gradient Approximation Made Simple". *Phys. Rev. Lett.* **1998**, *80*, 890.

(37) Feynman, R. P. Forces in Molecules. *Phys. Rev.* **1939**, *56*, 340.

(38) Parlinski, K.; Li, Z.-Q.; Kawazoe, Y. First-Principles Determination of the Soft Mode in Cubic ZrO$_2$. *Phys. Rev. Lett.* **1997**, *78*, 4063.

(39) Oba, F.; Togo, A.; Tanaka, I.; Paier, J.; Kresse, G. Defect energetics in ZnO: A hybrid Hartree-Fock density functional study. *Phys. Rev. B: Condens. Matter Mater. Phys.* **2008**, *77*,

245202.

(40) Darby, S.; Mortimer-Jones, T. V.; Johnston, R. L.; Roberts, C. Theoretical study of Cu-Au nanoalloy clusters using a genetic algorithm. *J. Chem. Phys.* **2002**, *116*, 1536.

(41) Gao, G. P.; Zheng, F.; Pan, F.; Wang, L.-W. Theoretical investigation of 2D conductive microporous coordination polymers as Li-S battery cathode with ultrahigh energy density. *Adv. Energy Mater.* **2018**, *8*, 1801823.

(42) Şahin, H.; Cahangirov, S.; Topsakal, M.; Bekaroglu, E.; Akturk, E.; Senger, R. T.; Ciraci, S. Monolayer honeycomb structures of group-IV elements and III-V binary compounds: First-principles calculations. *Phys. Rev. B: Condens. Matter Mater. Phys.* **2009**, *80*, 155453.

(43) Çakır, D.; Kecik, D.; Sahin, H.; Durgun, E.; Peeters, F. M. Realization of a p-n junction in a single layer boron-phosphide. *Phys. Chem. Chem. Phys.* **2015**, *17*, 13013.

(44) Zeng, B.; Li, M.; Zhang, X.; Yi, Y.; Fu, L.; Long, M. First-Principles Prediction of the Electronic Structure and Carrier Mobility in Hexagonal Boron Phosphide Sheet and Nanoribbons. *J. Phys. Chem. C* **2016**, *120*, 25037.

(45) Yu, T.-T.; Gao, P.-F.; Zhang, Y.; Zhang, S.-L. Boron-phosphide monolayer as a potential anchoring material for lithium-sulfur batteries: a first-principles study. *Appl. Surf. Sci.* **2019**, *486*, 281.

(46) Zhang, Y.; Wu, Z.-F.; Gao, P.-F.; Zhang, S.-L.; Wen, Y.-H. Could Borophene Be Used as a Promising Anode Material for High-Performance Lithium Ion Battery? *ACS Appl. Mater. Interfaces* **2016**, *8*, 22175.

(47) Zhang, Y.; Wu, Z.-F.; Gao, P.-F.; Fang, D.-Q.; Zhang, E.-H.; Zhang, S.-L. Structural, elastic, electronic, and optical properties of the tricycle-like phosphorene. *Phys. Chem. Chem. Phys.* **2017**, *19*, 2245.

(48) Wu, Z.-F.; Gao, P.-F.; Guo, L.; Kang, J.; Fang, D.-Q.; Zhang, Y.; Xia, M.-G.; Zhang, S.-L.; Wen, Y.-H. Robust indirect band gap and anisotropy of optical absorption in B-doped phosphorene. *Phys. Chem. Chem. Phys.* **2017**, *19*, 31796.

(49) Yang, L. M.; Bačić, V.; Popov, I. A.; Boldyrev, A. I.; Heine, T.; Frauenheim, T.; Ganz, E. Two-Dimensional $Cu_2Si$ Monolayer with Planar Hexacoordinate Copper and Silicon Bonding. *J. Am. Chem. Soc.* **2015**, *137*, 2757.

(50) Zhang, H.; Li, Y.; Hou, J.; Tu, K.; Chen, Z. $FeB_6$ Monolayers: The Graphene-like Material

with Hypercoordinate Transition Metal. *J. Am. Chem. Soc.* **2016**, *138*, 5644.

(51) Yu, T.; Zhao, Z.; Liu, L.; Zhang, S.; Xu, H.; Yang, G. C. TiC$_3$ Monolayer with High Specific Capacity for Sodium-Ion Batteries. *J. Am. Chem. Soc.* **2018**, *140*, 5962.

(52) Mouhat, F.; Coudert, F.-X. Necessary and Sufficient Elastic Stability Conditions in Various Crystal Systems. *Phys. Rev. B: Condens. Matter Mater. Phys.* **2014**, *90*, 224104.

(53) Cadelano, E.; Palla, P. L.; Giordano, S.; Colombo, L. Elastic Properties of Hydrogenated Graphene. *Phys. Rev. B: Condens. Matter Mater. Phys.* **2010**, *82*, 235414.

(54) Lee, C.; Wei, X.; Kysar, J. W.; Hone, J. Measurement of the Elastic Properties and Intrinsic Strength of Monolayer Graphene. *Science* **2008**, *321*, 385.

(55) Zhang, Y.; Wu, Z.-F.; Gao, P.-F.; Fang, D.-Q.; Zhang, E.-H.; Zhang, S.-L. Strain-tunable electronic and optical properties of BC$_3$ monolayer. *RSC Adv.* **2018**, *8*, 1686.

(56) Topsakal, M.; Cahangirov, S.; Ciraci, S. The response of mechanical and electronic properties of graphane to the elastic strain. *Appl. Phys. Lett.* **2010**, *96*, 091912.

(57) Ding, Y.; Wang, Y. Density Functional Theory Study of the Silicene-Like SiX and XSi$_3$ (X = B, C, N, Al, P) Honeycomb Lattices: The Various Buckled Structures and Versatile Electronic Properties. *J. Phys. Chem. C* **2013**, *117*, 18266.

(58) Choi, S.-M.; Jhi, S.-H.; Son, Y.-W. Effects of strain on electronic properties of grapheme. *Phys. Rev. B: Condens. Matter Mater. Phys* **2010**, *81*, 081407.

(59) Rakshit, B.; Mahadevan, P. Absence of rippling in graphene under biaxial tensile strain. *Phys. Rev. B: Condens. Matter Mater. Phys* **2010**, *82*, 153407.

**TOC Graphic**

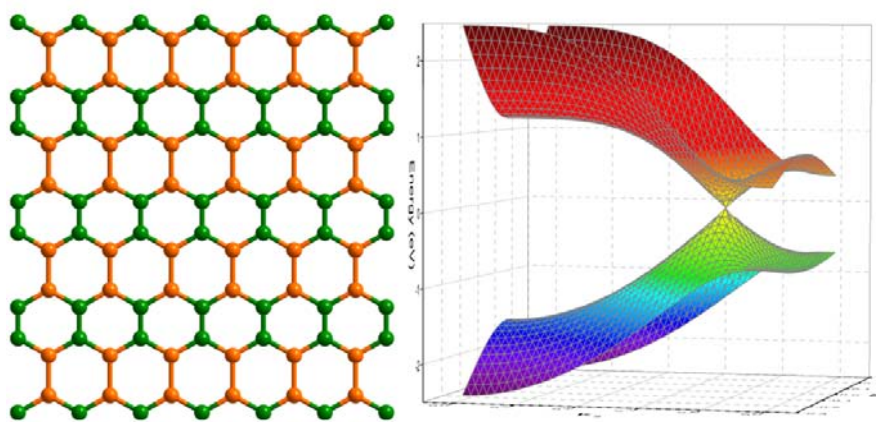

BP monolayer exhibits a robust Dirac cone with high Fermi velocity.